\documentclass[sigconf]{acmart}

\usepackage[utf8]{inputenc}
\usepackage{hyperref}
\usepackage{algorithm}
\usepackage{algpseudocode}
\usepackage{todonotes}
\usepackage{booktabs} 
\usepackage{tabularx}
\usepackage{graphicx}
\usepackage[caption=false,font=footnotesize]{subfig}
\usepackage{stfloats}
\usepackage{url}

\newcommand{\norm}[1]{\left\lVert #1 \right\rVert}
\newcommand{\abs}[1]{\left\lvert #1 \right\rvert}
\newcommand\Tstrut{\rule{0pt}{2.6ex}}
\newcommand\Bstrut{\rule[-0.9ex]{0pt}{0pt}}
\newcommand\ie{i.e.,\ }
\newcommand\eg{for example\ }

\newcolumntype{R}{>{\raggedleft\arraybackslash}X}
\newcommand{\revision}[1]{#1}

\setcopyright{rightsretained}

\begin{document}
\title[Parallel Algorithm for Approximate Calculation of Inverse $p$-th Roots of Sparse Matrices]{A Massively Parallel Algorithm for the Approximate Calculation of Inverse $p$-th Roots of Large Sparse Matrices}

\author{Michael Lass}
\orcid{0000-0002-5708-7632}
\affiliation{%
  \institution{Department of Computer Science, Paderborn University}
  \streetaddress{Warburger Str. 100}
  \postcode{33098}
  \city{Paderborn}
  \country{Germany}
}
\email{michael.lass@uni-paderborn.de}

\author{Stephan Mohr}
\orcid{0000-0003-2510-5805}
\affiliation{%
  \institution{Barcelona Supercomputing Center}
  \streetaddress{Carrer de Jordi Girona, 29}
  \postcode{08034}
  \city{Barcelona}
  \country{Spain}
}
\email{stephan.mohr@bsc.es}

\author{Hendrik Wiebeler}
\affiliation{%
  \institution{Department of Chemistry, Paderborn University}
  \streetaddress{Warburger Str. 100}
  \postcode{33098}
  \city{Paderborn}
  \country{Germany}
}
\email{wie@mail.uni-paderborn.de}

\author{Thomas D. Kühne}
\affiliation{%
  \institution{Department of Chemistry, Paderborn University}
  \streetaddress{Warburger Str. 100}
  \postcode{33098}
  \city{Paderborn}
  \country{Germany}
}
\email{tdkuehne@mail.uni-paderborn.de}

\author{Christian Plessl}
\affiliation{%
  \institution{Department of Computer Science, Paderborn University}
  \streetaddress{Warburger Str. 100}
  \postcode{33098}
  \city{Paderborn}
  \country{Germany}
}
\email{christian.plessl@uni-paderborn.de}


\begin{abstract}
We present the \emph{submatrix method}, a highly parallelizable method for the
approximate calculation of inverse $p$-th roots of large sparse symmetric
matrices which are required in different scientific applications.
We follow the idea of Approximate Computing, allowing imprecision in the final
result in order to be able to utilize the sparsity of the input matrix and to
allow massively parallel execution. For an $n\times n$ matrix, the proposed
algorithm allows to distribute the calculations over $n$ nodes with only little
communication overhead. The approximate result matrix exhibits the same sparsity
pattern as the input matrix, allowing for efficient reuse of allocated data
structures.

We evaluate the algorithm with respect to the error that it introduces into
calculated results, as well as its performance and scalability. We demonstrate
that the error is relatively limited for well-conditioned matrices \revision{and
that results are still valuable for error-resilient applications like
preconditioning even for ill-conditioned matrices}. We discuss the execution
time and scaling of the algorithm on a theoretical level and present a
distributed implementation of the algorithm using MPI and OpenMP. We demonstrate
the scalability of this implementation by running it on a high-performance
compute cluster comprised of 1024 CPU cores, showing a speedup of
\revision{$665\times$} compared to single-threaded execution.
\end{abstract}

\maketitle

\section{Introduction}

Inverse matrices and inverse $p$-th roots, \ie $A^{-1/p}$ for a given matrix
$A$, are important for various applications in the area of scientific computing.
Examples are preconditioning, solving systems of linear equations and linear
least squares problems, non-linear optimization, Kalman filtering and solving
generalized eigenvalue problems, in particular to solve Schrödinger and Maxwell
equations.

In\label{sec:relwork:linscaling} many of these applications, the involved
matrices are very large, containing billions of entries, and also sparse. For
increasing problem sizes, it becomes more and more important to exploit this
sparsity to save both computational effort and memory resources. However, the
inverse and inverse $p$-th roots of a sparse matrix are typically not sparse
anymore, which makes exploitation of the sparsity difficult. \revision{The novel
approach proposed in this work is to only compute an approximate solution for
the inverse p-th roots of large matrices and enforcing that the result has the
same sparsity pattern as the input matrix. At the same time the method allows
massive parallelization of the required computations.} Hence, the proposed
method adopts the idea of \emph{Approximate Computing}, which denotes
the concept of sacrificing accuracy of computation results in order to increase
the efficiency of these computations~\cite{KlavikMalossiBekasEtAl2014}. \revision{We call the method proposed in this work the \emph{submatrix method}.}

There are multiple applications where approximate solutions for inverse $p$-th
roots are of use. One of these applications is preconditioning, where efficient
computations can be more important than accuracy. Another particularly important
application area are electronic structure methods to approximately solve the
electronic Schrödinger equation~\cite{Kohn1999}. Interestingly it has been shown
that, for large system sizes, the representing matrices become eventually all
sparse~\cite{Goedecker1999}. Underlying this is the concept of ``nearsightedness
of electronic matter'', which states that, at fixed chemical potential, the
electronic density depends just locally on the external
potential~\cite{prodan2005}. Consequently, with increasing system size, the
number of nonzero elements in matrices used to simulate these systems only
increases linearly, \ie the density decreases also linearly as the system size
increases. Specialized solver libraries such as CheSS~\cite{Mohr2017} exploit
this behavior \revision{but still require processing large sparse matrices}. The
method proposed in this work is especially suitable in these applications since
it exploits the increasing sparsity of the matrices \revision{and allows scaling
the parallelism with the matrix size}. Since linear-scaling electronic structure
methods entail approximation in practice~\cite{Richters2014}, it can be
sufficient to calculate an approximation for the inverse $p$-th root of the
involved matrices.

The remainder of this work is structured as follows. In
Section~\ref{sec:relwork} we give a brief overview of related work in the field
of matrix inversion, calculation of $p$-th roots and parallelization of these
operations. In Section~\ref{sec:algo} we present the submatrix method in detail.
Afterwards, we discuss the error that is introduced by using the submatrix
method in Section~\ref{sec:error} and its complexity and scalability in
Section~\ref{sec:complexity}. Finally, we present an implementation of the
proposed method using MPI~\cite{MPI} and OpenMP~\cite{OpenMP} in
Section~\ref{sec:eval} and use it to evaluate the scalability of the method in
practice using a large compute cluster. We conclude in
Section~\ref{sec:conclusion}.

\section{Foundations and Related Work}\label{sec:relwork}
Due to the importance of matrix inversions and inverse $p$-th roots
and the computational complexity of these operations, there is a large variety
of ressources on different numerical methods and on efficient implementation of
these methods. We want to give a brief overview of commonly used algorithms,
ready-to-use implementations and related work on parallelization of the required
calculations. However, an exhaustive coverage of available methods is
outside the scope of this work.

The inverse $X$ of a matrix $A$ fulfils the equation
\begin{align}
AX=I,
\end{align}
where $I$ is the identity matrix. $X$ can therefore be determined by solving
this equation using Gaussian elimination or by calculating and using an LU or
LUP decomposition of $A$. For symmetric matrices, one can compute the
singular value decomposition (SVD), and invert all singular values. A different
approach is to reduce the problem of determining the inverse of an $n\times n$
input matrix to calculating the inverse of two $n/2\times n/2$ matrices and
performing several matrix-matrix multiplications. Following this idea
recursively, inverting a matrix can be reduced to performing only matrix-matrix
multiplications~\cite[Ch. 28.2]{cormen}. Lastly, iterative methods
can be used to find the inverse of a matrix, such as the Newton-Schulz iteration
scheme~\cite{schulz1933}.

In literature, several approaches can be found to parallelize matrix inversion
or calculation of LU and SV decompositions. For example,
Van der Stappen et al.~\cite{vanderStappen1993} present an algorithm for
parallel calculation of the LU decomposition on a mesh network of transputers
where each processor holds a part of the matrix. Shen~\cite{Shen2006} evaluates
techniques for LU decomposition distributed over nodes that are connected via
slow message passing.
Dongarra et al.~\cite{Dongarra2011} demonstrate an optimized implementation of
matrix inversion on a single multicore node, focusing on the minimization of
synchronization between the different processing cores.
There are also algorithms specialized on specific applications, such as the one
described by Lin et al.~\cite{Lin2011} which can be used in 2D electronic
structure calculations to only calculate selected parts of the inverse of a
sparse matrix. For parallel calculation of the SVD, Berry et
al.~\cite{berry2006} provide an extensive overview of parallelizable methods.

For the calculation of the $p$-th root $A^{1/p}$ of a matrix $A$, a commonly
used algorithm is the one described by Higham and
Lin~\cite{higham2011,higham2013}. For the calculation of inverse $p$-th roots,
\ie $A^{-1/p}$, there are also iterative methods available, such as the ones
described by Bini et al.~\cite{bini2005algorithms} and Richters et
al.~\cite{richters17_arxiv}, which reduce the problem to repeated matrix-matrix
multiplications. For symmetric matrices, the solution can again also be computed
by building the SVD of the input matrix and applying the operation of interest
to all singular values of the matrix. \revision{For very large sparse matrices,
calculation of the SVD is typically avoided. Instead methods are used where only
the largest eigenvalues of the matrix are calculated. Iterative methods to do so
are the Lanczos algorithm~\cite{lanczos1950} and the Arnoldi
iteration~\cite{arnoldi1951}.}

Implementations for the calculation of the LU and SV decompositions and matrix
inversion are part of LAPACK~\cite{Angerson1990}, a popular software library for
numerical linear algebra. For solving large sparse systems and calculation of
singular values, ARPACK~\cite{arpack} is a well known library \revision{which is
based on the Arnoldi iteration}. There exist different implementations of these
libraries, as well as bindings for many different programming languages. With
ScaLAPACK~\cite{Choi1996} and P\_ARPACK~\cite{p_arpack}, there exist extensions
of these libraries targeting parallel execution on distributed memory systems
using MPI for message passing.

\section{Algorithm Description}\label{sec:algo}
\begin{figure*}[t]
  \includegraphics[width=\textwidth]{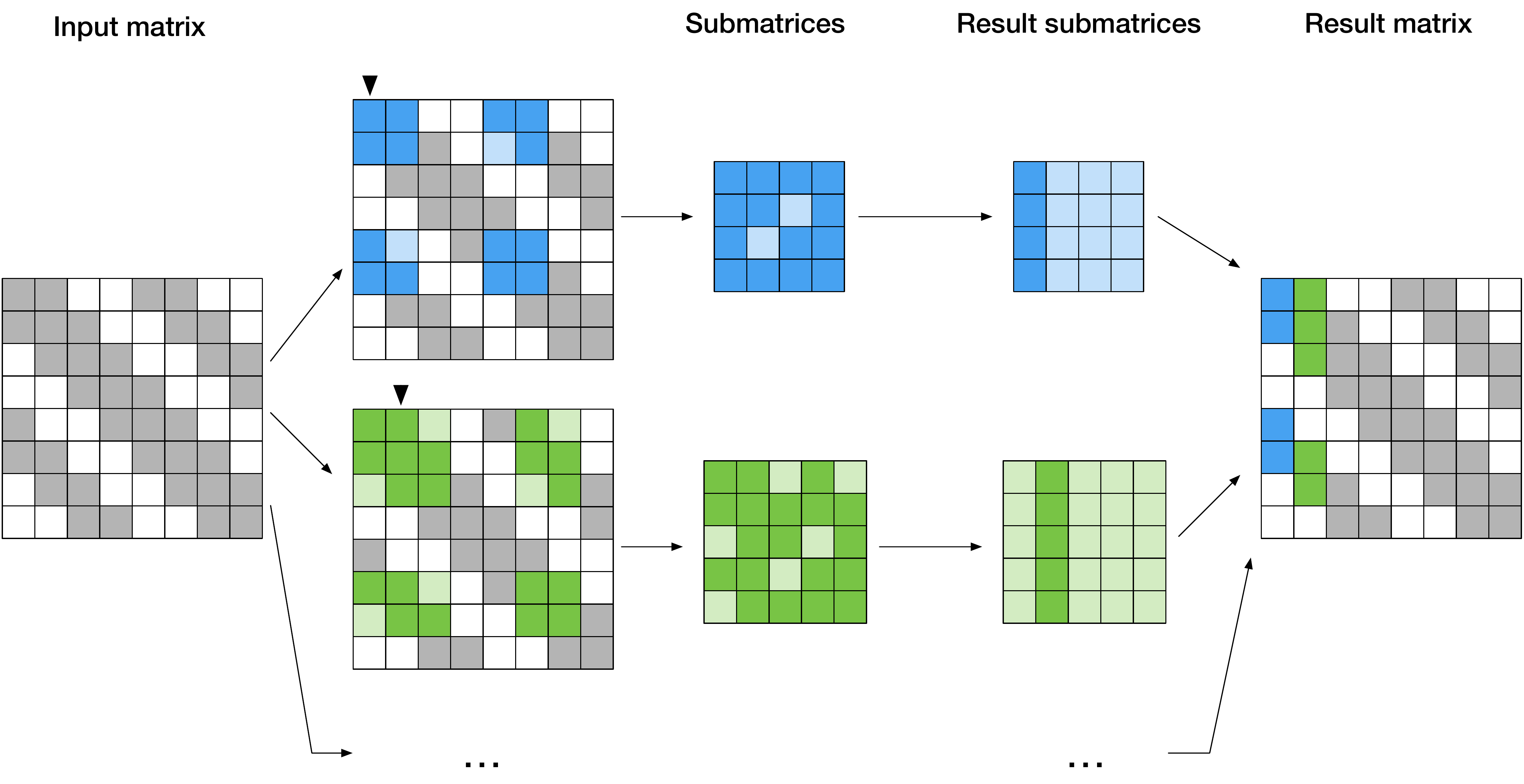}
  \caption{\label{fig:overview}Overview of the submatrix method}
\end{figure*}

The fundamental concept of the proposed submatrix method is to divide the large,
sparse input matrix into several smaller and dense submatrices and to apply the
desired operation, such as inversion or calculation of inverse $p$-th roots, to
all of these submatrices instead of the original matrix. Afterwards the results
of these operations are used to construct an approximate result matrix. The
overall procedure is shown in Figure~\ref{fig:overview}. In the following, we
describe the main steps of this method, in particular, building the submatrices
and assembling the final result matrix, in detail. Note that, although we only
discuss a column-based approach for building the submatrices, the method can as
well be applied in a row-based manner since we are dealing with symmetric
matrices. For ease of reading, we will often only mention the inversion of
matrices in the remainder of this section, while the same also holds for the
calculation of inverse $p$-th roots.

\subsection{Building the submatrices}
The process of building the submatrices is shown in Algorithm~\ref{lst:divide}.
To simplify comprehension of the algorithm, all matrices have a dense
representation in our pseudocode. As will be discussed in
Section~\ref{sec:implnotes}, in practice a sparse representation should be used
for the sparse input and output matrices.

\begin{algorithm}[tb]
  \begin{algorithmic}

    \State $n \gets$ number of rows/columns of input matrix
    \State $A[1\dots n][1\dots n] \gets$ input matrix

    \For {$j \gets 1\dots n$}
      \State $R\gets\emptyset$
      \For {$i\gets 1\dots n$}
        \If {$A[i][j]\neq 0$}
          \State $R\gets R \cup \{i\}$
        \EndIf
      \EndFor
      \State $m\gets R$.length()
      \For {$k\gets 1\dots m$}
        \For {$l\gets 1\dots m$}
          \State submatrices$[j][k][l] \gets A[R[k]][R[l]]$
        \EndFor
      \EndFor
      \State indices$[j] \gets R$ \Comment{required later on for result assembly}
    \EndFor

  \end{algorithmic}
  \caption{\label{lst:divide}Construction of submatrices}
\end{algorithm}

To construct the $j$-th submatrix, the $j$-th column of the input matrix $A$ is
evaluated. We determine the set $R$ of row indices $i$ for which $A_{i,j}\neq
0$. The submatrix is then constructed by taking all values $A_{x,y}$ from the
input matrix where \mbox{$x,y\in R$}. For an input matrix of size $n\times n$ we
obtain a set of $n$ submatrices. The size of each submatrix is determined by the
number of nonzero elements in the corresponding column of the input matrix.

\subsection{Performing submatrix operations}
For all of the submatrices, we now perform the operation which should originally
be performed on the input matrix, \ie we either invert all submatrices or
calculate their inverse $p$-th roots. Note that the method and implementation
for these submatrix operations can be freely selected and this choice is
entirely orthogonal to the submatrix method described in this work.

\subsection{Assembling the result matrix}
After having applied the matrix operation of interest to each submatrix, we have
$n$ result submatrices. From these result submatrices we assemble an approximate
solution $X$ for the whole matrix. This procedure is shown in
Algorithm~\ref{lst:combine}.

\begin{algorithm}[tb]
  \begin{algorithmic}

    \State $n \gets$ number of rows/columns of input matrix
    \State indices$[1\dots n] \gets$ from submatrix generation stage
    \State submatrices$[1\dots n][1\dots ?][1\dots ?] \gets$ result matrices

    \State $X\gets$ zeros($n\times n$)
    \For{$j\gets 1\dots n$}
      \State $R\gets$ indices$[j]$
      \State $m\gets R$.length()
      \For{$i\gets 1\dots m$}
        \State $X[R[i]][j] \gets$ submatrices$[j][i][R$.indexof$(j)]$
      \EndFor
    \EndFor

  \end{algorithmic}
  \caption{\label{lst:combine}Assembly of result matrix}
\end{algorithm}

Similar to the construction of the submatrices, the $j$-th column of the final
result matrix is determined by the $j$-th result submatrix. We take the values
from the column of the result submatrix which was originally filled with values
from the $j$-th column of the input matrix, and copy them back to their original
position in the original matrix.

\subsection{Implementation notes}\label{sec:implnotes}
Although the inverse of a sparse matrix is typically not sparse, the approximate
solution provided by the submatrix method exhibits the exact same sparsity
pattern as the input matrix. This allows for efficient implementation of the
method based on matrices in the compressed sparse column (CSC) format which
consists of a value array (\texttt{val}), a list of row indices
(\texttt{row\_ind}) and a list of column pointers (\texttt{col\_ptr}). In
particular, the result assembly stage can be implemented by concatenating the
corresponding columns of all result submatrices to obtain the value array
\texttt{val} for the approximate result matrix. \texttt{row\_ind} and
\texttt{col\_ptr} from the input matrix can be reused for the output matrix
without any changes. If the method is applied in a row-based manner, the same
holds for matrices in the compressed sparse row (CSR) format.

\section{Applicability and Approximation Error}\label{sec:error}
Evidently, the result obtained by the submatrix method is only an approximation
of the correct result. Whether this result is still of use for an application
depends on three aspects: Is the application able to deal with results that
contain a certain error, how large can this error be to be still tolerable and
how large is the error introduced into results by using the submatrix method. In
this section, we first characterize the error based on random sparse matrices
and afterwards apply the submatrix method to \revision{two application scenarios
and show that using the submatrix method to process real-world matrices in these
application yields good results.}
Finally we discuss means to influence the approximation error.

\subsection{Error for random input matrices}\label{sec:error:generic}
To get an impression about the error introduced for arbitrary symmetric
positive-definite matrices, we generate random matrices $A$ using the
\texttt{sprandsym}\footnote{\texttt{sprandsym(size,density,1/condition,kind)} with \texttt{kind}=1} function in Matlab. 
\revision{This allows us to sweep over different sizes $n$, densities $d$ and condition numbers $\kappa$ and assess the influence of these matrix properties onto the eror.}
For each set of these
parameters, we generate ten different matrices. For all of these matrices we
then use the submatrix method to obtain an approximate solution $X$ for the
inverse $p$-th root $A^{-1/p}$. To assess the error of these results, we
calculate the spectral norm of the residuals
\begin{align}
\norm{R}_2 &= \norm{X^p A - I}_2.
\end{align}
Since for a precise solution it should hold that $X^p A = I$, $R$ is a good
indicator for the introduced error. We choose the spectral norm of $R$ as a
metric because in contrast to other matrix norms like the Frobenius norm it is
relatively invariant of the matrix size. The spectral norm of a matrix is
defined as
\begin{align}
\norm{M}_2 &= \sqrt{\lambda_\text{max}(M^{H} M)},
\end{align}
where $\lambda_\text{max}(M)$ denotes the largest eigenvalue of $M$ and $M^{H}$
is the Hermitian transpose of $M$.

\begin{figure}[tb]
  \includegraphics[width=\columnwidth,trim=2cm 0.5cm 0 1cm]{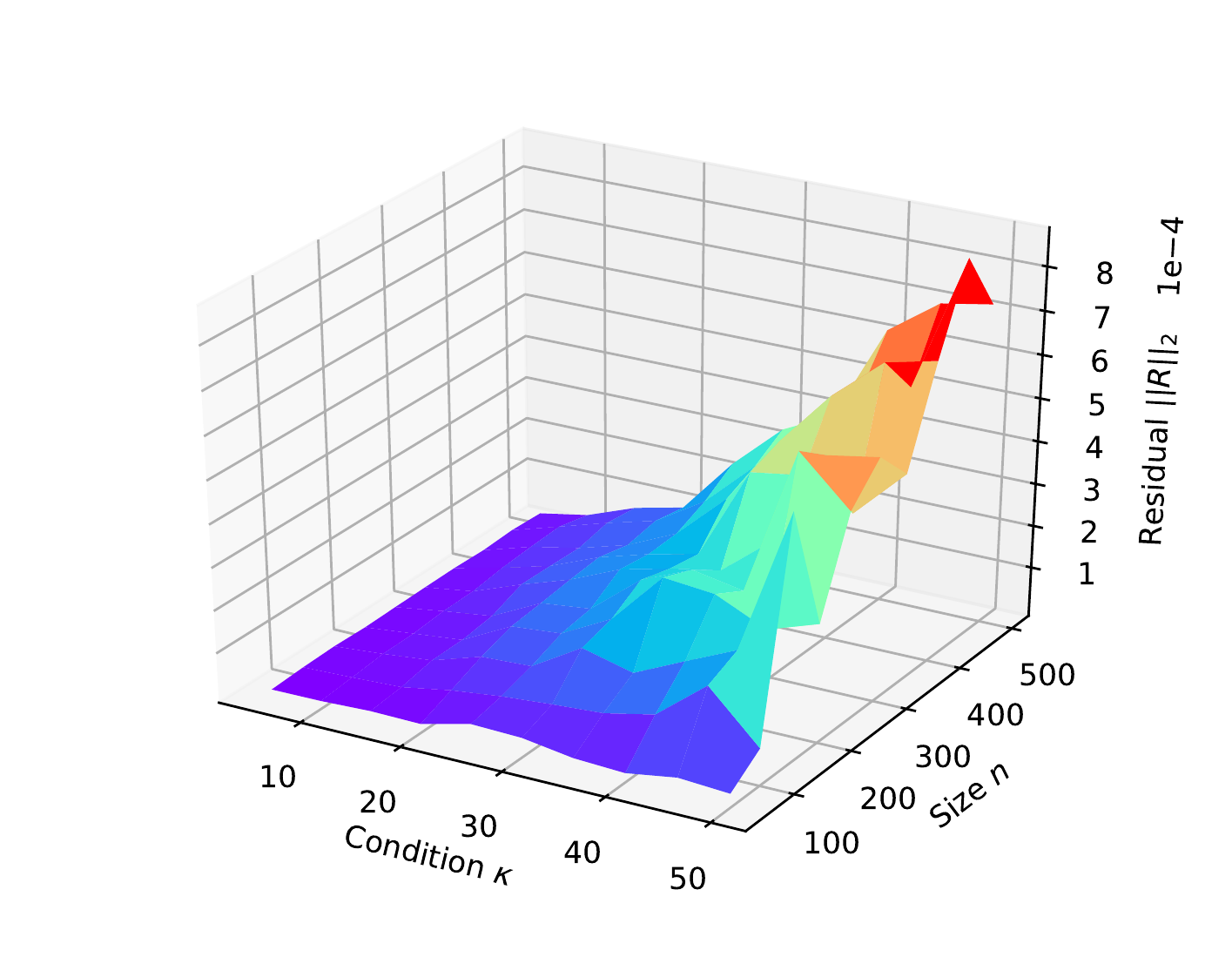}
  \caption{\label{fig:surface}Residual for approximately calculated inverse of random matrices using submatrix method, for different sizes and condition numbers.}
\end{figure}

Our initial evaluation has not shown a significant influence of the density of
the randomly generated matrices onto the precision of the result. We therefore
neglect this parameter in the evaluation presented here, focus on matrices with
density $d=0.05$ and discuss the influence of the size and the condition number
of the matrices. Figure~\ref{fig:surface} shows the relationship between these
matrix properties and the calculated residual for $p=1$. It shows
that the error increases for matrices with larger size and larger condition
numbers. For small matrices, the error stays relatively low even for higher
condition numbers. Similarly, for well-conditioned matrices, the error stays low
even for large matrices.

\begin{figure}[bt]
  \includegraphics[width=\columnwidth]{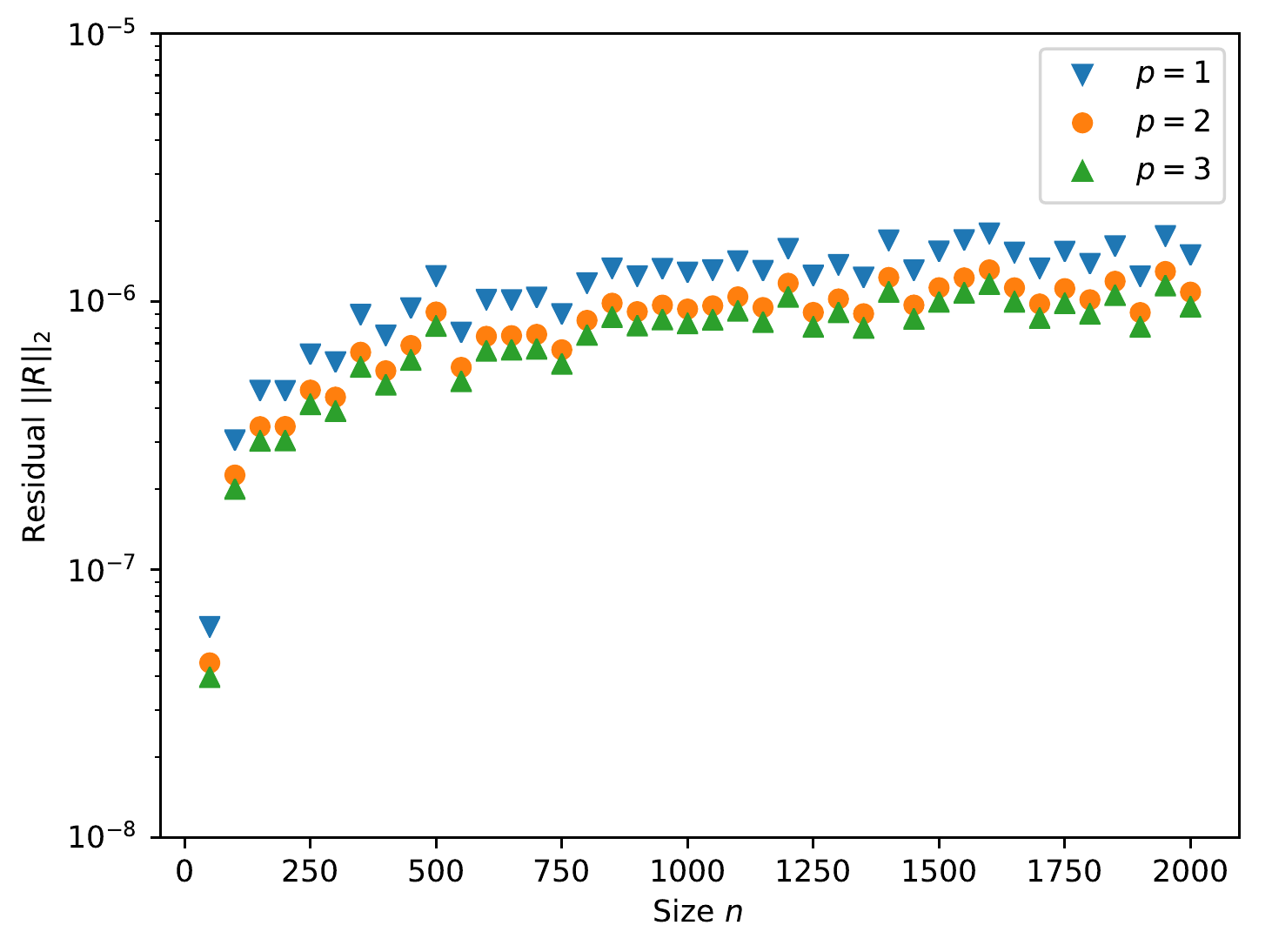}
  \caption{\label{fig:error-size}Residual for approximately calculated inverse of random matrices with $\kappa=2$ using submatrix method, in relation to size of input matrix.}
\end{figure}

To demonstrate the latter, we now focus on \revision{well-conditioned} matrices with $\kappa=2$
and $d=0.05$, varying only their size. Results are shown in
Figure~\ref{fig:error-size}. It shows that for a fixed condition number, the
error introduced by using the submatrix method is limited even when further
increasing the matrix size. As shown, this not only holds for calculating the
inverse of a matrix but also for calculating inverse $p$-th roots where $p>1$.

\subsection{Applicability to other matrix operations}
Throughout this work we describe the submatrix method as a method to calculate
inverse $p$-th roots of matrices. \revision{In fact, the method can be applied
to similar matrix operations as well, such as the calculation of positive $p$-th
roots where the error behaves very similar as discussed in Section~\ref{sec:error:generic}.}
%
%
\revision{In contrast to the case of inverse $p$-th roots, the
residual is calculated as $R = X^p A^{-1} - I$, which requires a matrix inversion itself.}
Hence, the residual cannot be calculated efficiently at runtime to assess the
error of an approximately computed result. Inverse $p$-th roots are therefore,
and due to their variety of target applications, presented as the main target in
this work.

\subsection{Application within electronic structure codes}\label{sec:application}
In Section~\ref{sec:error:generic} we demonstrated that using the
submatrix method \revision{for well-conditioned matrices} yields results very similar to a precisely calculated
solution. However,
whether errors are acceptable in an application depends on the the kind of
matrices used in the application and the effect that small deviations have on
the final result. In this section, we show a specific application of the
submatrix method and demonstrate its limited influence on the final result.

We choose one specific time-consuming kernel of effective
single-particle theory, the orthogonalization of a set of $n$ non-orthogonal
basis functions, and  assess the impact of using the submatrix method within
this kernel on  the band-structure energy $E_{\text{BS}}$, which corresponds to
the sum of eigenvalues of the Hamilton matrix $H$. For that purpose, we first
compute the so-called  overlap matrix $S\in \mathbb{R}^{n\times n}$, where
$S_{i,j} = \langle \varphi_i | \varphi_j \rangle$ and $\varphi_i$ are the $n$
non-orthogonal basis functions spanning the Hilbert space. Specifically, we
consider overlap matrices of bulk liquid water systems  for different numbers of
molecules and hence different dimension of the the overlap matrix, as computed
using a Daubechies Wavelet-based density functional theory (DFT)
code~\cite{mohr2014}.

\begin{figure}
  \centering
  \includegraphics[width=.7\columnwidth]{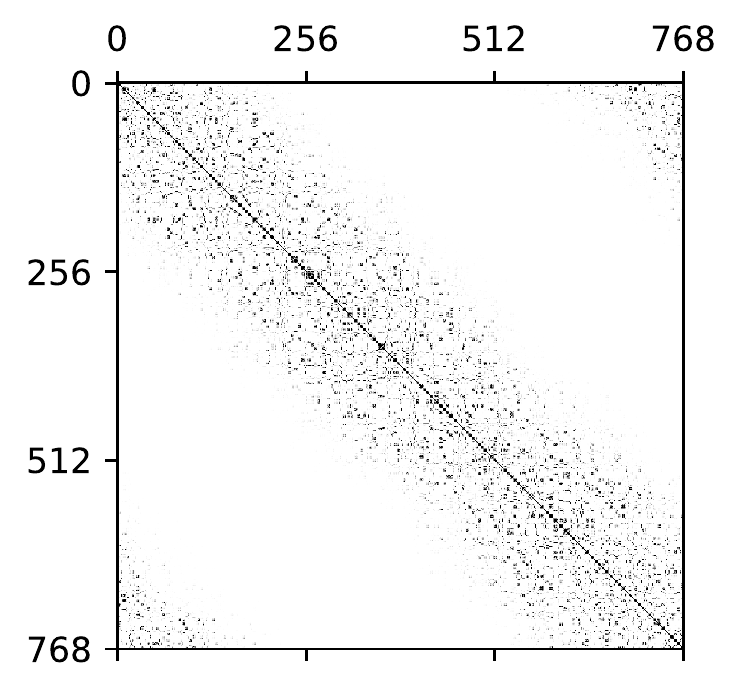}
  \caption{Structure of an examined overlap matrix.}
\label{fig:overlapMatrix}
\end{figure}

Figure~\ref{fig:overlapMatrix} shows one such overlap matrix for
a system of 128 H\textsubscript{2}O molecules. The matrix has a banded structure
and in this case a density of $d=0.25$ and a condition number of around
$\kappa=1.5$. For increasing system sizes, the density decreases lineraly with
$n$, \eg $d=0.12$ for $n=1536$, $d=0.06$ for $n=3072$ etc., while condition
numbers remain approximately constant at around $\kappa=1.5$.

In addition to the overlap matrices $S$, we also extract the density
matrix $P$ as well as the Hamilton matrix $H$ from our Wavelet-based DFT code.
This allows us to calculate the band-structure energy as
\begin{align}
  E_\text{BS} = \text{tr}(PH).
\end{align}
To orthogonalize the Hamilton matrix, we calculate
\begin{align}
  H_\text{ortho} = HS^{-1},\label{eq:ortho}
\end{align}
where $S^{-1}$ is computed using the submatrix method. From this, we again
calculate the band-structure energy as
\begin{align}
  E_\text{BS}^\text{sm} = \text{tr}(SPH_\text{ortho}),
\end{align}
and evaluate the relative error caused by the approximative inversion of $S$ as
\begin{align}
  \Delta E_\text{rel} = \abs{\frac{   E_\text{BS} - E_\text{BS}^\text{sm}   }{   E_\text{BS}   }}.
\end{align}

\begin{table}[t]
  \centering
  \caption{\label{tab:energy}Influence of using the submatrix method for orthogonalization of basis functions onto electronic band structure energy calculations.}
  \begin{tabularx}{\columnwidth}{rRRR}
    \hline\hline
    \Tstrut\Bstrut\textbf{Matrix size} $n$ & $E_\text{BS}$ & $E_\text{BS}^\text{sm}$ & $\Delta E_\text{rel}$\\
    \hline
      768 &   -372.83597 &   -372.83600 & 6.96$\times 10^{-8}$\Tstrut\\
     1536 &   -747.13928 &   -747.13933 & 7.60$\times 10^{-8}$\\
     3072 &  -1492.25282 &  -1492.25297 & 1.01$\times 10^{-7}$\\
     6144 &  -2986.25656 &  -2986.25683 & 8.76$\times 10^{-8}$\\
    12288 &  -5976.31525 &  -5976.31576 & 8.64$\times 10^{-8}$\\
    24576 & -11951.19504 & -11951.19598 & 7.85$\times 10^{-8}$\Bstrut\\
    \hline\hline
  \end{tabularx}
\end{table}

Results are shown in Table~\ref{tab:energy}. For all evaluated matrix
sizes, the relative error caused by using the submatrix method for
orthogonalization is rather small and throughout below or around $10^{-7}$.

\revision{
The band structure of the considered overlap matrices may suggest that the low deviation between $E_\text{BS}$ and
$E_\text{BS}^\text{sm}$ comes from the similarity between $S$ and the identity
matrix $I$, and therefore $I$ could be used in Equation~(\ref{eq:ortho}) as an
approximation for $S^{-1}$. However, doing so leads to relative errors between
$1.15\times 10^{-2}$ and $1.21\times 10^{-2}$, \ie five orders of magnite higher
than when using the approximate inverse provided by the submatrix method.}

\subsection{\revision{Application as preconditioner}}\label{sec:preconditioning}
\revision{In the last section we showed an application where the input matrices are
well conditioned and the results provided by the submatrix method are good enough
for the application. Now we demonstrate using the submatrix method to process ill-conditioned matrices in order to obtain a
preconditioner that can be used to iteratively solve systems of linear equations.}

\begin{table}[t]
  \centering
  \caption{\label{tab:preconditioning}\revision{Number of iterations required to solve Equation~(\ref{eq:lineq}) for different matrices $A$ using CG with different preconditioners.}}
  \begin{tabularx}{\columnwidth}{lRRRRR}
    \hline\hline
    \Tstrut\Bstrut\textbf{Matrix} & $n$ & $\kappa$ & \textbf{None} & \textbf{SM} & \textbf{ILU(0)} \\
    \hline

    1138\_bus & 1138 & 8.5$\times10^{06}$ & 2120 & 151 & 139\Tstrut\\
    bcsstk08 & 1074 & 2.6$\times10^{07}$ & --- & 41 & 27\\
    bcsstk09 & 1083 & 9.5$\times10^{03}$ & 194 & 56 & ---\\
    bcsstk10 & 1086 & 5.2$\times10^{05}$ & --- & 85 & 182\\
    bcsstk11 & 1473 & 2.2$\times10^{08}$ & --- & 273 & 477\\
    bcsstk12 & 1473 & 2.2$\times10^{08}$ & --- & 273 & 477\\
    bcsstk13 & 2003 & 1.1$\times10^{10}$ & --- & 409 & ---\\
    bcsstk14 & 1806 & 1.2$\times10^{10}$ & --- & 54 & 262\\
    bcsstk15 & 3948 & 6.5$\times10^{09}$ & --- & 177 & 591\\
    bcsstk16 & 4884 & 4.9$\times10^{09}$ & 464 & 32 & 35\\
    bcsstk21 & 3600 & 1.8$\times10^{07}$ & --- & 224 & ---\\
    bcsstk23 & 3134 & 2.7$\times10^{12}$ & --- & 1269 & ---\\
    bcsstk24 & 3562 & 2.0$\times10^{11}$ & --- & 300 & 244\\
    bcsstk26 & 1922 & 1.7$\times10^{08}$ & --- & 325 & 337\\
    bcsstk27 & 1224 & 2.4$\times10^{04}$ & 907 & 66 & 19\\
    bcsstk28 & 4410 & 9.5$\times10^{08}$ & --- & 668 & 755\\
    bcsstm12 & 1473 & 6.3$\times10^{05}$ & 2790 & 7 & 12\\
    Chem97ZtZ & 2541 & 2.5$\times10^{02}$ & 86 & 10 & 1\\
    crystm01 & 4875 & 2.3$\times10^{02}$ & 70 & 8 & 2\\
    ex10hs & 2548 & 5.5$\times10^{11}$ & --- & --- & ---\\
    ex10 & 2410 & 9.1$\times10^{11}$ & --- & --- & ---\\
    ex13 & 2568 & 1.1$\times10^{15}$ & --- & --- & ---\\
    ex33 & 1733 & 7.0$\times10^{12}$ & --- & 1052 & ---\\
    ex3 & 1821 & 1.7$\times10^{10}$ & --- & --- & ---\\
    ex9 & 3363 & 1.2$\times10^{13}$ & --- & --- & ---\\
    mhd3200b & 3200 & 1.6$\times10^{13}$ & --- & 6 & 3\\
    mhd4800b & 4800 & 8.2$\times10^{13}$ & --- & 6 & 2\\
    msc01050 & 1050 & 4.6$\times10^{15}$ & --- & --- & ---\\
    msc01440 & 1440 & 3.3$\times10^{06}$ & --- & 89 & 155\\
    msc04515 & 4515 & 2.3$\times10^{06}$ & 4411 & 357 & ---\\
    nasa1824 & 1824 & 1.9$\times10^{06}$ & --- & 275 & 264\\
    nasa2146 & 2146 & 1.7$\times10^{03}$ & 282 & 67 & 12\\
    nasa2910 & 2910 & 6.0$\times10^{06}$ & --- & 282 & 760\\
    nasa4704 & 4704 & 4.2$\times10^{07}$ & --- & 1100 & 570\\
    plat1919 & 1919 & 1.2$\times10^{17}$ & --- & --- & ---\\
    plbuckle & 1282 & 1.3$\times10^{06}$ & 1965 & 76 & 69\\
    sts4098 & 4098 & 2.2$\times10^{08}$ & --- & 67 & 119\\
    Trefethen\_2000 & 2000 & 1.6$\times10^{04}$ & 435 & 6 & 5\Bstrut\\

    \hline\hline
  \end{tabularx}
\end{table}

\revision{To demonstrate this application scenario, we obtain sparse, symmetric,
positive definite matrices from the SuiteSparse matrix library~\cite{florida}.
We select all matrices $A$ with size $1000\leq n \leq 5000$ that fulfill these
requirements. For these matrices we solve the system
\begin{align}
  Ax=b,\quad b=[1,1,\dots,1]^T\label{eq:lineq}
\end{align}
using the Conjugate Gradient (CG) method. We set the threshold for the residual
to $10^{-6}$ and limit the number of iterations by $2n$.
Table~\ref{tab:preconditioning} shows the number of iterations required for CG
to converge towards a solution.}
\revision{For preconditioning, we use the submatrix method to obtain
\begin{align}
  K \approx A^{-1/2}.
\end{align}
Instead of solving Equation~(\ref{eq:lineq}), we now solve the system given by
\begin{align}
  K^T A K y = K^T b
\end{align}
using the CG method. The solution $x$ for Equation~(\ref{eq:lineq}) can then be
computed as
\begin{align}
  x = K y.
\end{align}
Again, results are shown in Table~\ref{tab:preconditioning}. For comparison, we
also include the number of iterations required when using an
ILU(0)\footnote{incomplete LU decomposition with zero fill-in} preconditioner.
The results show that using the submatrix method for preconditioning is not only
competitive to the use of ILU(0) but enables CG to converge in more of the
cases.}

\subsection{Controlling the approximation error}
We have demonstrated the use of the submatrix method in \revision{two
applications that can highly benefit from the speedup and the additional
parallelism and still yields good results}. However, there may be applications
that are less tolerant to errors but still can benefit from using the submatrix
method.

If an application requires a lower error than what is provided by the
solution calculated using the submatrix method, an iterative method like those
described by Bini et al.~\cite{bini2005algorithms} and Richters et
al.~\cite{richters17_arxiv} can be used to refine the solution obtained from the
submatrix method. The result obtained by using the submatrix method then acts as
an initial guess for these iterative methods. While we validated that such a
refinement of a solution generated by the submatrix method works in principle
and converges within very few iterations, a detailed evaluation of combining the
submatrix method with iterative methods remains for future work.

In the contrary case, if the application has a particularly high
resiliency against errors in the inverse matrix, the submatrix method can also
be combined with other approximation techniques to achieve further performance
gains. Since using the submatrix method is orthogonal to the implementation of
the operations performed on the single submatrices, these submatrix calculations
can be performed in an approximate manner as well. Using an iterative method,
precision can be scaled by the number of iterations. Additionally, calculations
can be performed using low precision arithmetic or other Approximate Computing
techniques~\cite{lass17_esl}.

\section{Complexity and Scalibility}\label{sec:complexity}
We now want to discuss the time complexity and scalability of the submatrix
method and show that, although for an $n\times n$ matrix $n$ submatrices need to
be processed, it can still provide a significant reduction in time required for
determining a matrix inverse or its inverse $p$-th root.

Note that the considerations in this section only hold if the density
of the input matrix shows in each of its columns (or rows). An obvious
counterexample are arrowhead matrices, where using the submatrix method cannot
provide any speedup since the first submatrix has the same size as the original
input matrix.

\subsection{Single-threaded scenario}
We first want to discuss the general complexity of matrix inversion, both using
conventional methods and using our proposed submatrix method. While from a
theoretical standpoint, inversion of matrices is not harder than multiplication,
and therefore $\mathcal{O}(n^{2.81})$ using Strassen's algorithm~\cite[pp. 79,
829]{cormen}, or even $\mathcal{O}(n^{2.373})$ using Coppersmith and Winograd's
algorithm~\cite{LeGall2014}, in practice methods such as Gaussian elimination or
building and using the LU decomposition for inversion which have time complexity
$\mathcal{O}(n^3)$  are commonly used. In the following, we define $I(n)$ as the
time required for a precise matrix inversion, abstracting from a concrete
implementation.

For a sparse $n\times n$ input matrix, using the submatrix method requires
performing $n$ matrix inversions for smaller but dense matrices. To be more
efficient in a single-threaded application scenario, these submatrices have to
be significantly smaller than the original input matrix. In the
following, we assume a uniformly filled, sparse input matrix. Let $d$ be the
density of this matrix, then the average size $m\times m$ of the submatrices is
determined by:
\begin{align}
m &= d\cdot n.
\end{align}
If the density $d$ is small enough, such that
\begin{align}
n\cdot I(d\cdot n) &< I(n),\label{eq:breakeven}
\end{align}
then the submatrix method has lower run time than a precise inversion, even in a
single-threaded scenario.

We now want to determine, by what rate the density $d$ has to decrease so that
for increasing matrix sizes the asymptotic run time does not grow faster than
using conventional methods for matrix inversion. We therefore assume that matrix
inversion has at least time complexity $n^2$, \ie $I(n) = \Omega(n^2)$. To
fulfill Equation~(\ref{eq:breakeven}), it then needs to hold that for
sufficiently large $n$
\begin{align}
n \cdot {(d \cdot n)}^2 &< n^2\nonumber\\
d^2 &< n^{-1}\nonumber\\
d &< n^{-0.5}.
\end{align}
From this we can deduce the following asymptotic relation:
\begin{align}
d = \mathcal{O}(n^{-0.5}) &\Rightarrow S(n,d) = \mathcal{O}(I(n)),\label{eq:o-notation}
\end{align}
where $S(n,d)$ is the time required to calculate an approximate inverse using
the submatrix method. If $d$ decreases faster than with rate $n^{-0.5}$, then
$S(n,d)$ increases slower than $I(n)$ for larger $n$. Using methods where
$I(n)=\Omega(n^3)$ further relaxes the requirements on $d$, in that
$d=\mathcal{O}(n^{-1/3})$ suffices to fulfill Equation~(\ref{eq:breakeven}) for
large $n$.

Note that we neglect the time required for building the submatrices and
assembling the final result matrix, as their influence on execution time is
negligible compared to the involved matrix inversions in asymptotic
considerations.

\subsection{Parallel execution of submatrix operations}
Although using the submatrix method can reduce execution time even in a
single-threaded environment for very sparse matrices, its strength is to allow
massively parallel execution. All submatrix operations are entirely independent
from each other such that the inversion of an $n\times n$ matrix can be
distributed over $n$ compute nodes. Each compute node can construct its own
submatrix from the input matrix. The final result matrix has to be assembled on
a single node but as described in Section~\ref{sec:implnotes}, this step
consists of a simple concatenation of $n$ arrays. Communication between nodes is
only required for initial data distribution and for the final collection of all
results. Provided that $n$ compute nodes can be used for execution of the
algorithm, a speedup is already achievable if all submatrices are significantly
smaller than the original input matrix, \ie each column of the original matrix
contains a significant fraction of zero-elements.

\subsection{Application to Electronic Structure Methods}\label{sec:complexity:linscaling}
In Section~\ref{sec:relwork:linscaling}, we motivated our method with the
principal of linear scaling techniques in density functional theory, based on
the nearsightedness of electronic matter. With respect to the matrices for which
an inverse or an inverse $p$-th root needs to be calculated, this means that
while for growing systems the total number of matrix elements increases
quadratically, the density of the matrix decreases linearly with $n^{-1}$.
Consequently, the number of nonzero elements in the matrix increases only
linearly with $n$.

Based on this fact, the submatrix method is particularly suitable for solving
these problems. In particular, since
\begin{align}
  n^{-1} < n^{-0.5},
\end{align}
the density of matrices decreases faster than required in
Equation~(\ref{eq:o-notation}). From that it follows that the asymptotic run
time of the submatrix method in a single-threaded environment is limited by that
of a precise inversion for the appliations discussed here.

Again, the strong advantage of the submatrix method is the possibility of
parallel execution on many compute nodes. In the case of linear scaling methods
in density functional theory, this means that for growing systems the execution
can be parallelized onto more and more nodes while the size of the single
submatrices stays constant. As long as the number of compute nodes can be scaled
with $n$ as well, the overall execution time can even be held constant.

\section{Performance Evaluation}\label{sec:eval}
To evaluate the performance and scalability of the proposed method, we built a
distributed implementation using MPI and OpenMP\@. We run this implementation on
a compute cluster comprised of 65 nodes. Each node features two Intel Xeon
E5-2670 CPUs with a total of 16 CPU cores and 64~GByte of memory. All nodes are
connected via 40~Gbit/s QDR InfiniBand. We use one node as a control node,
leaving the remaining 64 nodes with a total of 1024 CPU cores for handling the
workload.
In the following, we first describe details of our implementation, and
then present results obtained from running our implementation on our compute
cluster.

\subsection{Details of our Implementation}
Our implementation makes use of Intel MPI~\cite{IntelMPI} to distribute work
over a large number of compute nodes and to collect all results in order to
build up the final result matrix. The MPI rank 0, in the following called main
process, reads the input matrix stored in CSC format from persistent storage
into memory. Metadata such as an identifyer for the matrix, as well as its total
size and number of nonzero elements, are then sent via \texttt{MPI\_Bcast} to
all nodes.

\subsubsection{Data distribution and work assignment}
There are different possible ways to make the input matrix available to all
other MPI ranks, which we call worker processes in the following. In principal,
it would be sufficient to send single submatrices to the workers which then
perform the inversion. In this case, all submatrices would have to be
constructed within the main process, which would clearly present a bottleneck.
Instead, we make the whole input matrix available to all worker processes which
then autonomously construct their submatrices. In our environment all systems
have access to a shared file system which allows all processes to read in the
input matrix from persistent storage. Since this scenario cannot generally be
assumed, we additionally implemented distribution of data via
\texttt{MPI\_Bcast} to all worker processes. We found that, for the data we use
in our evaluation, \revision{both variants provide comparable performance}.

Assuming we have $w$ workers, each worker needs to process \mbox{$x=n/w$}
submatrices. In our implementation, each worker processes a contiguous set of
submatrices, \ie the worker with rank $k$ is responsible for submatrices
$(k-1)x$ to $kx-1$. Therefore, depending on its rank and the total number of
ranks, each worker process can determine autonomously, which of the submatrices
it has to process.\footnote{Note that if the number of worker processes does not
divide the size of the matrix, some workers need to process one additional
submatrix.} It builds the submatrix according to Algorithm~\ref{lst:divide} and
calls the LAPACK functions \texttt{dgetrf} to obtain an LU decomposition and
\texttt{dgetri} to calculate the inverse of the submatrix. In our evaluation we
use Intel MKL~\cite{IntelMKL} as a highly optimized implementation for these
LAPACK routines. After inversion, the worker selects the section of the result
matrix which is relevant for the final result matrix and stores it in a buffer.
Since the main process just needs to concatenate these buffers to create the
final result matrix, a single call to \texttt{MPI\_Gatherv} is sufficient to
perform data collection and assembly after all submatrices have been processed.

\subsubsection{Multi-threading using OpenMP}
The described implementation already allows to distribute the load over many
nodes and CPU cores. To utilize multiple CPU cores on a single node,
multiple MPI ranks could be placed on a node, or multiple cores could be used for
processing a single submatrix by using a multi-threaded LAPACK implementation.
Having multiple MPI ranks on the same node comes at the cost of data duplication
in memory and overall increased MPI communication load. Using multiple cores for
a single submatrix operation has shown to provide lower speedup than additional
parallelization of submatrix operations.

In our implementation, we therefore use OpenMP to process $c$ submatrices in
parallel on a node featuring $c$ CPU cores. We do so by calling all submatrix
operations within an OpenMP parallel for loop:

\begin{center}
\texttt{\small \#pragma omp parallel for schedule(dynamic)}
\end{center}

To allow OpenMP to fully utilize the available CPU cores, we explicitly disable
the multi-threading functionality provided by Intel MKL by calling
\texttt{mkl\_set\_num\_threads(1)}.

\subsubsection{Limitations of our Implementation}\label{sec:limitations}
As described, each worker process is responsible for a contigous set of
submatrices and all workers are responsible for the same number ($\pm 1$) of
submatrices. This can lead to workload imbalance between the different workers,
if the input matrix exhibits a pattern such that certain sets of columns contain
significantly more or significantly fewer nonzero values than other sets of
columns. It is important to note that this is a limitation of our implementation
and not a conceptual issue of the proposed submatrix method. In practice, there
are different ways to deal with this issue in order to create an optimized
implementation that does not exhibit this load imbalance:

\emph{Shuffling input matrices:} To balance the load between all worker
processes, the mapping between submatrices and workers can be shuffled randomly.
Clusters of full colums which result in larger submatrices would then not be
assigned to a single worker but distributed over all workers. This can, for
example, be implemented using a pseudo-random but deterministic permutation, so
that each worker can still autonomously determine the submatrices it is
responsible for.

In our implementation, each worker concatenates the results of its submatrix
operations in a buffer which is then sent as a whole to the main process. The
main process therefore only needs to collect and concatenate $w$ arrays for $w$
worker processes. If submatrices are shuffled, collection and cocatenation of
$n$ arrays would be required in the main process instead. \revision{Apart from
this, there is no additional computational effort required for this load
balancing technique.}

\emph{Dynamic work scheduling:} Instead of assigning a fixed set of submatrices
to a worker process, work can be scheduled dynamically. Each worker could
request work packages from the main process using MPI\@. The size of these work
packages can be chosen in the range from one single submatrix up to $n/w$
submatrices in order to trade off load balancing and additional communication
effort. \revision{Concepts similar to OpenMP's \emph{guided} scheduling could also be implemented to
minimize scheduling overhead.}
Note that in our implementation, we already use dynamic work scheduling for the
parallel processing of multiple submatrices on a single node by using OpenMP's
\emph{dynamic} scheduler.

\subsubsection{Availability}
Our prototype implementation as well as scripts used in our evaluation are
published under MIT license and can be found on
\url{https://github.com/pc2/SubmatrixMethod}.

\subsection{Results}

\subsubsection{Scalability for increasing number of CPU cores}\label{sec:res:scalability}
\begin{figure}
  \includegraphics[width=\columnwidth]{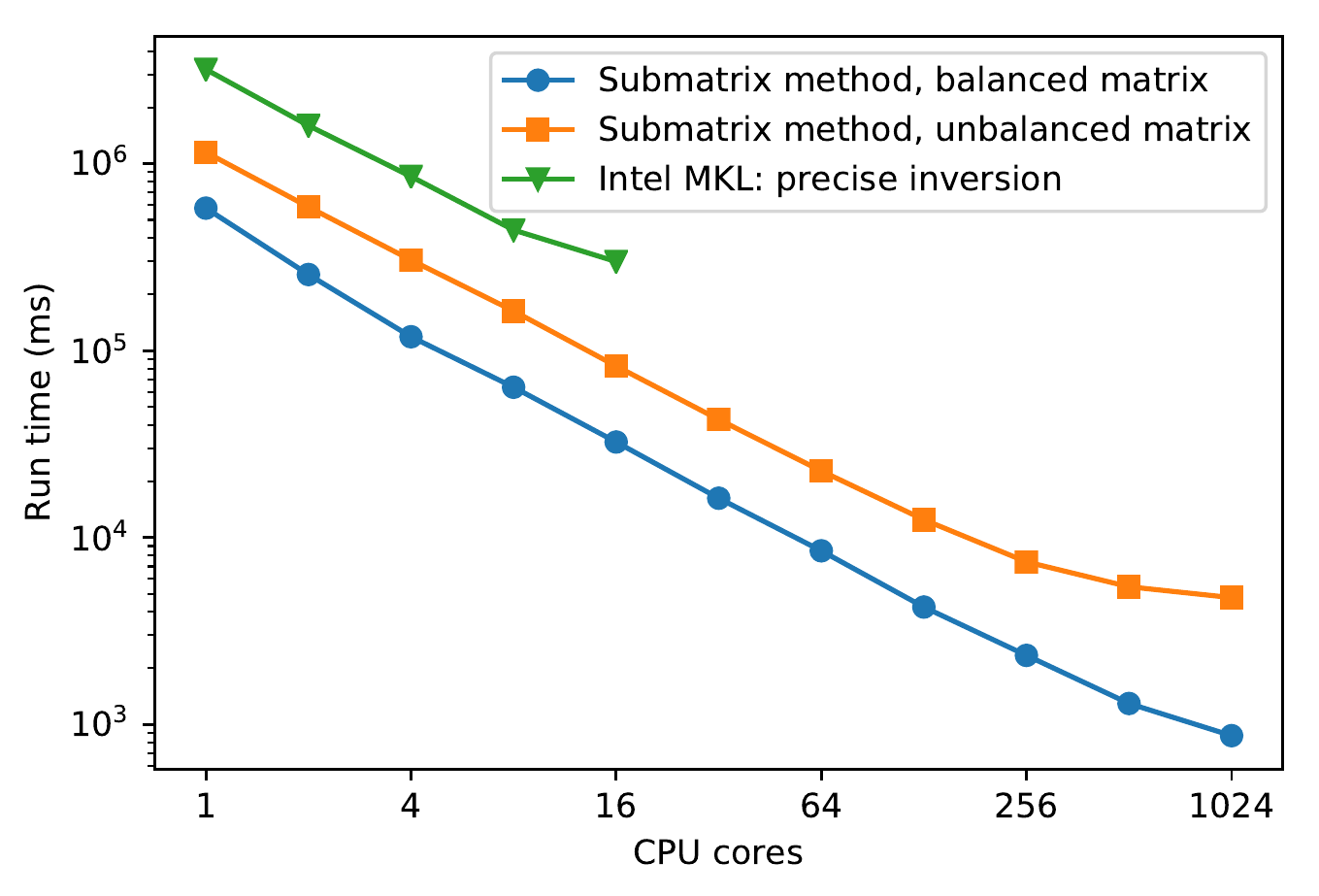}
  \caption{\label{fig:scaling-d1}\revision{Scalability of the submatrix method for a random matrix of size $n=32768$ and density $d=0.01$.}}
\end{figure}

We use our implementation of the submatrix method to calculate an approximate
inverse of multiple random matrices with size $n=32768$, condition number
$\kappa =2$ and density $d=0.01$. We vary the number of utilized CPU cores in
the range from 1 to 1024 and measure the total wall clock time required to
obtain a result. \revision{We consider a set of balanced matrices whose columns have
roughly the same number of nonzero
elements\footnote{generated using \texttt{sprandsym(size,density,1/condition,kind)} with
\texttt{kind}=2} and a set of unbalanced matrices which exhibit visible patterns
in the distribution of
values\footnote{generated using \texttt{sprandsym(size,density,1/condition,kind)} with
\texttt{kind}=1}.}
Results
are shown in Table~\ref{tab:scaling-d1} and Figure~\ref{fig:scaling-d1}. \revision{As a reference}, we also show the time required for a precise matrix inversion using
Intel MKL's implementation of the \texttt{dgetrf} and \texttt{dgetri} routines,
utilizing up to 16 CPU cores on a single node.

\begin{table}[b]
  \centering
  \caption{\label{tab:scaling-d1}\revision{Time in ms required for inversion of a matrix with size $n=32768$ and density $d=0.01$ using the submatrix method.}}
  \begin{tabularx}{\columnwidth}{rRRRR}
    \hline\hline
    \Tstrut & \multicolumn{2}{c}{\textbf{Balanced matrix}} & \multicolumn{2}{c}{\textbf{Unbalanced matrix}}\\
    \Bstrut\textbf{Cores} & \textbf{Wall time} & \textbf{Speedup} & \textbf{Wall time} & \textbf{Speedup}\\
    \hline
         1 & 578,140 & 1.0 & 1,150,366 & 1.0\Tstrut\\
         2 & 255,081 & 2.3 & 586,778 & 2.0\\
         4 & 118,534 & 4.9 & 304,941 & 3.8\\
         8 & 63,644 & 9.1 & 162,792 & 7.1\\
        16 & 32,405 & 17.8 & 82,571 & 13.9\\
        32 & 16,216 & 35.7 & 42,760 & 26.9\\
        64 & 8,485 & 68.1 & 22,692 & 50.7\\
       128 & 4,242 & 136.3 & 12,447 & 92.4\\
       256 & 2,339 & 247.2 & 7,402 & 155.4\\
       512 & 1,293 & 447.1 & 5,447 & 211.2\\
      1024 & 870 & 664.5 & 4,765 & 241.4\Bstrut\\
    \hline\hline
  \end{tabularx}
\end{table}

\revision{The results show that the submatrix method overall scales well
over a large number of processors. Comparing the data for
balanced and unbalanced matrices, two distinct effects can be observed:
\begin{enumerate}
\item Even for a low number of CPU cores, the submatrix method performs better for balanced matrices. The reason for this is that if some columns contain significantly more nonzero elements than others, the resulting submatrices are larger in size and the time to process them increases cubically with their size.
\item For the imbalanced matrices in our scenario, the curve starts to flatten at around 256 cores and scaling beyond 512 cores provides diminishing returns. The reason for this is that the number of submatrices per worker becomes small enough such that load imbalance between workers has an increasing effect. This effect could be countered by implementing some form of load balancing, as discussed in Section~\ref{sec:limitations}.
\end{enumerate}
For over 512 cores, even for the balanced matrices in our scenario the additional speedup is limited. This is caused by the overall short runtime of the algorithm and therefore increased influence of communication time (around 32\%).
}

On a single node, Intel MKL as well nearly scales linearly with the number of
CPU cores. Only for 16 cores there is a slight efficiency drop\revision{, likely caused by the NUMA architecture of our compute nodes}. Using ScaLAPACK
to distribute execution of the utilized library functions over multiple nodes
may allow to further increase the number of CPU cores. However, due to
increasing communication overhead, the potential for scaling is limited in this
case. Related work that uses ScaLAPACK for matrix inversion, describes
decreasing performance for execution on more than 64 CPU
cores~\cite{Kolberg2008}.

\subsubsection{Run time for growing matrices}
We now evaluate how the total execution time develops for increasing matrix
sizes, given a fixed number of CPU cores.
We consider two different scenarios: a fixed density of $d=0.01$ \revision{and matrix sizes ranging from $2^{11}$ to $2^{18}$} and a density that decreases
linearly with $n$ as encountered in applications like electronic structure
methods. For the latter, we set $d=0.16 \cdot 1024 / n$ \revision{and consider sizes from $2^{10}$ to $2^{20}$}.

\begin{figure}
  \includegraphics[width=\columnwidth]{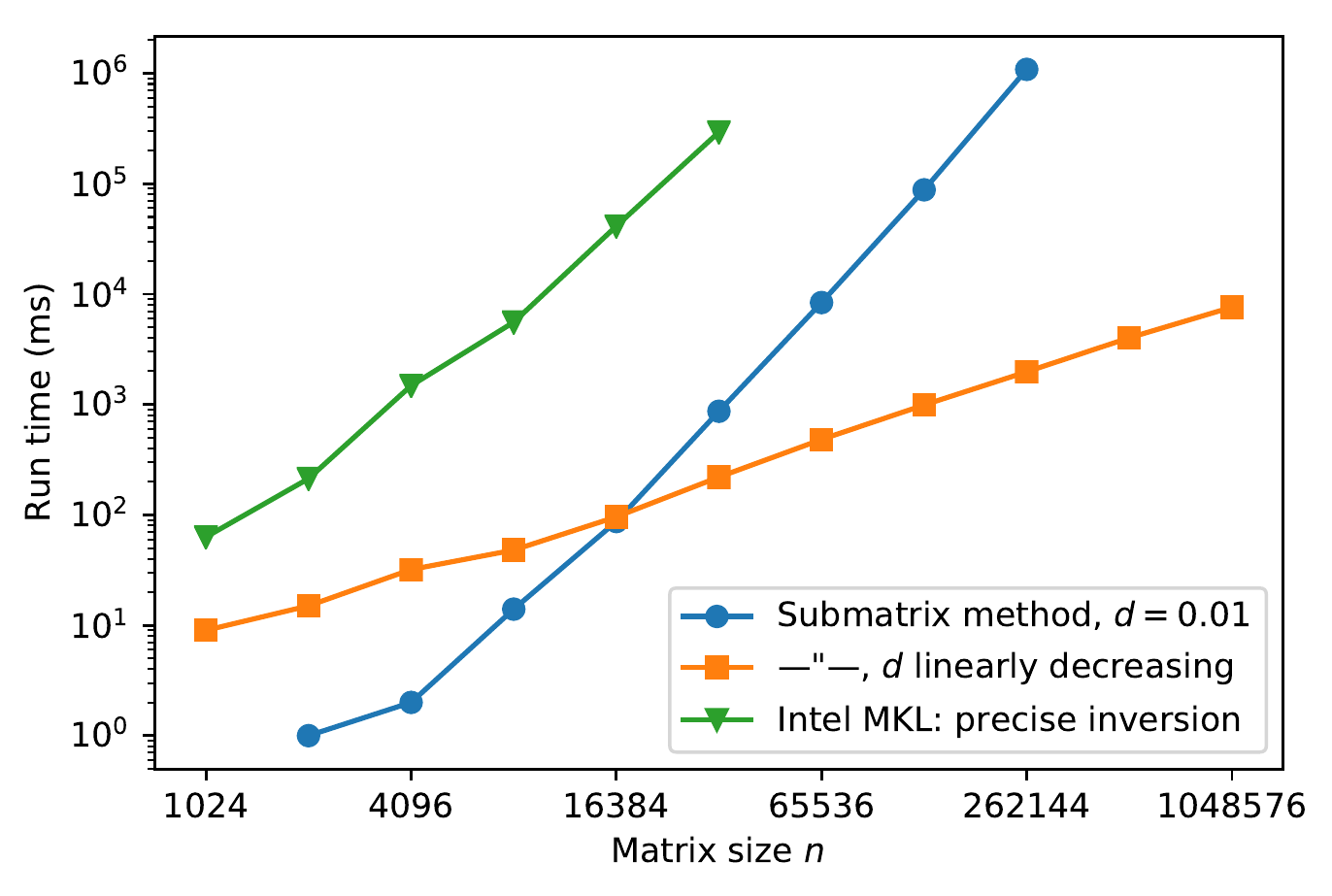}
  \caption{\label{fig:matrix-sizes}\revision{Required time for inversion of a matrix using the submatrix method (1024 cores) and Intel MKL (16 cores).}}
\end{figure}

The results of this evaluation are shown in Table~\ref{tab:matrix-sizes} and
Figure~\ref{fig:matrix-sizes}.
\revision{In the table we also show the fraction of the total wall clock time spent on communication and the fraction of compute time spent on building the submatrices. Note that the assembly of the result matrix is performed implicitly by \texttt{MPI\_Gatherv} and therefore accounted as communication time.}
It clearly shows that for matrices
with linearly decreasing density, the required run time \revision{only increases linearly with the matrix size}, as expected based on the discussion in
Section~\ref{sec:complexity:linscaling}. Combining this result with the
possibility for linear performance scaling with the number of CPU cores, the run
time can be held constant by increasing the number of cores with $n$ for growing
matrices. \revision{The data also shows that for increasing size of the submatrices, as shown in the upper half of Table~\ref{tab:matrix-sizes}, the overhead required for communication and for building the submatrices decreases. For fixed-size submatrices, the overhead stays relatively constant. Note that times for communication fluctuate in our measurements due to shared usage of the underlying InfiniBand network.}

\begin{table}[t]
  \centering
  \caption{\label{tab:matrix-sizes}\revision{Time in ms required for inversion of a matrix using the submatrix method on 1024 cores.}}
  \begin{tabularx}{\columnwidth}{rRRRR}
    \hline\hline
    \Tstrut\Bstrut\textbf{Size} & \textbf{Density} & \textbf{Wall time} & \textbf{MPI Comm.} & \textbf{Submat. Constr.}\\
    \hline
        2,048 & $1.0\times 10^{-2}$ & 1 & --- & ---\Tstrut\\
        4,096 & $1.0\times 10^{-2}$ & 2 & --- & ---\\
        8,192 & $1.0\times 10^{-2}$ & 14 & 57.1\% & 58.8\%\\
       16,384 & $1.0\times 10^{-2}$ & 87 & 39.1\% & 60.5\%\\
       32,768 & $1.0\times 10^{-2}$ & 868 & 31.7\% & 57.8\%\\
       65,536 & $1.0\times 10^{-2}$ & 8,380 & 13.5\% & 56.8\%\\
      131,072 & $1.0\times 10^{-2}$ & 87,977 & 5.0\% & 47.0\%\\
      262,144 & $1.0\times 10^{-2}$ & 1,085,176 & 1.5\% & 36.8\%\Bstrut\\
    \hline
      1,024 & $1.6\times 10^{-1}$ & 9 & 22.2\% & 61.3\%\Tstrut\\
      2,048 & $8.0\times 10^{-2}$ & 15 & 26.7\% & 61.0\%\\
      4,096 & $4.0\times 10^{-2}$ & 32 & 37.5\% & 60.5\%\\
      8,192 & $2.0\times 10^{-2}$ & 48 & 33.3\% & 60.3\%\\
      16,384 & $1.0\times 10^{-2}$ & 96 & 38.5\% & 60.4\%\\
      32,768 & $5.0\times 10^{-3}$ & 220 & 51.7\% & 60.9\%\\
      65,536 & $2.5\times 10^{-3}$ & 482 & 54.4\% & 61.4\%\\
      131,072 & $1.3\times 10^{-3}$ & 990 & 54.7\% & 62.2\%\\
      262,144 & $6.3\times 10^{-4}$ & 1977 & 55.2\% & 63.0\%\\
      524,288 & $3.1\times 10^{-4}$ & 4020 & 56.8\% & 63.7\%\\
      1,048,576 & $1.6\times 10^{-4}$ & 7609 & 49.9\% & 64.0\%\Bstrut\\
    \hline\hline
  \end{tabularx}
\end{table}

\section{Conclusion}\label{sec:conclusion}
In this work we presented the submatrix method, which can be used to calculate
an approximate inverse of matrices, as well as inverse $p$-th roots. Following
the idea of Approximate Computing, it allows the result to deviate from an
exactly calculated solution in order to utilize the sparsity of the input matrix
and to allow massively parallel execution of the involved calculations. For an
$n\times n$ matrix, the workload can be distributed over $n$ nodes.

A particularly well suited application for the submatrix method are electronic
structure methods in density functional theory. In these applications, for
growing matrices
their density decreases linearly at the same time. In this
case, the submatrix method exhibits a \revision{linear} increase in execution time for
growing systems. As long as the number of available CPU cores can be scaled with
the same rate, execution time can even be held constant.

We showed that the error introduced by using the submatrix method is limited for
well-conditioned input matrices \revision{and demonstrated its use for
preconditioning of ill-conditioned matrices}. We discussed the scalability of
the algorithm both theoretically and in a practical evaluation on a large
compute cluster.

\revision{An emerging question for future work is how this method can be
combined with iterative methods to refine the solution or to increase efficiency
of the submatrix operations. Additionally, the dense nature of the submatrices
make the submatrix operations well suited for the use of accelerator hardware
which may be explored in future to develop high-performance implementations of
the submatrix operations.}

\begin{acks}
This project has received funding from the \grantsponsor{erc}{European Research
Council (ERC)}{https://erc.europa.eu/} under the European Union's Horizon 2020
research and innovation programme (grant agreement No~\grantnum{erc}{716142})
and from the \grantsponsor{dfg}{German Research Foundation
(DFG)}{http://www.dfg.de/en/index.jsp} under the project PerficienCC
(grant agreement No~\grantnum{dfg}{PL 595/2-1}). The evaluation was
performed on resources provided by the Paderborn Center for Parallel Computing
(PC\textsuperscript{2}).
\end{acks}

\bibliographystyle{ACM-Reference-Format}
\bibliography{lass18_pasc}


\begin{thebibliography}{32}


\ifx \showCODEN    \undefined \def \showCODEN     #1{\unskip}     \fi
\ifx \showDOI      \undefined \def \showDOI       #1{#1}\fi
\ifx \showISBNx    \undefined \def \showISBNx     #1{\unskip}     \fi
\ifx \showISBNxiii \undefined \def \showISBNxiii  #1{\unskip}     \fi
\ifx \showISSN     \undefined \def \showISSN      #1{\unskip}     \fi
\ifx \showLCCN     \undefined \def \showLCCN      #1{\unskip}     \fi
\ifx \shownote     \undefined \def \shownote      #1{#1}          \fi
\ifx \showarticletitle \undefined \def \showarticletitle #1{#1}   \fi
\ifx \showURL      \undefined \def \showURL       {\relax}        \fi
\providecommand\bibfield[2]{#2}
\providecommand\bibinfo[2]{#2}
\providecommand\natexlab[1]{#1}
\providecommand\showeprint[2][]{arXiv:#2}

\bibitem[\protect\citeauthoryear{Angerson, Bai, Dongarra, Greenbaum, McKenney,
  Croz, Hammarling, Demmel, Bischof, and Sorensen}{Angerson
  et~al\mbox{.}}{1990}]%
        {Angerson1990}
\bibfield{author}{\bibinfo{person}{E. Angerson}, \bibinfo{person}{Z. Bai},
  \bibinfo{person}{J. Dongarra}, \bibinfo{person}{A. Greenbaum},
  \bibinfo{person}{A. McKenney}, \bibinfo{person}{J.~Du Croz},
  \bibinfo{person}{S. Hammarling}, \bibinfo{person}{J. Demmel},
  \bibinfo{person}{C. Bischof}, {and} \bibinfo{person}{D. Sorensen}.}
  \bibinfo{year}{1990}\natexlab{}.
\newblock \showarticletitle{{LAPACK}: A portable linear algebra library for
  high-performance computers}. In \bibinfo{booktitle}{\emph{Proceedings
  SUPERCOMPUTING '90}}. \bibinfo{pages}{2--11}.
\newblock


\bibitem[\protect\citeauthoryear{Arnoldi}{Arnoldi}{1951}]%
        {arnoldi1951}
\bibfield{author}{\bibinfo{person}{W.~E. Arnoldi}.}
  \bibinfo{year}{1951}\natexlab{}.
\newblock \showarticletitle{The principle of minimized iteration in the
  solution of the matrix eigenvalue problem}.
\newblock \bibinfo{journal}{\emph{Quart. Appl. Math.}}  \bibinfo{volume}{9}
  (\bibinfo{year}{1951}), \bibinfo{pages}{17--29}.
\newblock


\bibitem[\protect\citeauthoryear{Berry, Mezher, Philippe, and Sameh}{Berry
  et~al\mbox{.}}{2006}]%
        {berry2006}
\bibfield{author}{\bibinfo{person}{Michael~W Berry}, \bibinfo{person}{Dani
  Mezher}, \bibinfo{person}{Bernard Philippe}, {and} \bibinfo{person}{Ahmed
  Sameh}.} \bibinfo{year}{2006}\natexlab{}.
\newblock \showarticletitle{Parallel algorithms for the singular value
  decomposition}.
\newblock \bibinfo{journal}{\emph{Statistics Textbooks and Monographs}}
  \bibinfo{volume}{184}, \bibinfo{number}{117} (\bibinfo{year}{2006}),
  \bibinfo{pages}{31}.
\newblock


\bibitem[\protect\citeauthoryear{Bini, Higham, and Meini}{Bini
  et~al\mbox{.}}{2005}]%
        {bini2005algorithms}
\bibfield{author}{\bibinfo{person}{D.~A. Bini}, \bibinfo{person}{N.~J. Higham},
  {and} \bibinfo{person}{B. Meini}.} \bibinfo{year}{2005}\natexlab{}.
\newblock \showarticletitle{Algorithms for the matrix pth root}.
\newblock \bibinfo{journal}{\emph{Numerical Algorithms}} \bibinfo{volume}{39},
  \bibinfo{number}{4} (\bibinfo{year}{2005}), \bibinfo{pages}{349--378}.
\newblock


\bibitem[\protect\citeauthoryear{Choi, Demmel, Dhillon, Dongarra, Ostrouchov,
  Petitet, Stanley, Walker, and Whaley}{Choi et~al\mbox{.}}{1996}]%
        {Choi1996}
\bibfield{author}{\bibinfo{person}{J. Choi}, \bibinfo{person}{J. Demmel},
  \bibinfo{person}{I. Dhillon}, \bibinfo{person}{J. Dongarra},
  \bibinfo{person}{S. Ostrouchov}, \bibinfo{person}{A. Petitet},
  \bibinfo{person}{K. Stanley}, \bibinfo{person}{D. Walker}, {and}
  \bibinfo{person}{R.~C. Whaley}.} \bibinfo{year}{1996}\natexlab{}.
\newblock \bibinfo{booktitle}{\emph{{ScaLAPACK}: A portable linear algebra
  library for distributed memory computers --- Design issues and performance}}.
\newblock \bibinfo{publisher}{Springer Berlin Heidelberg},
  \bibinfo{address}{Berlin, Heidelberg}, \bibinfo{pages}{95--106}.
\newblock
\showISBNx{978-3-540-49670-0}


\bibitem[\protect\citeauthoryear{Cormen, Leiserson, Rivest, and Stein}{Cormen
  et~al\mbox{.}}{2009}]%
        {cormen}
\bibfield{author}{\bibinfo{person}{T.H. Cormen}, \bibinfo{person}{C.E.
  Leiserson}, \bibinfo{person}{R.L. Rivest}, {and} \bibinfo{person}{C. Stein}.}
  \bibinfo{year}{2009}\natexlab{}.
\newblock \bibinfo{booktitle}{\emph{Introduction to Algorithms}
  (\bibinfo{edition}{3} ed.)}.
\newblock \bibinfo{publisher}{MIT Press}.
\newblock
\showISBNx{9780262533058}
\showLCCN{2009008593}


\bibitem[\protect\citeauthoryear{Dagum and Menon}{Dagum and Menon}{1998}]%
        {OpenMP}
\bibfield{author}{\bibinfo{person}{L. Dagum} {and} \bibinfo{person}{R. Menon}.}
  \bibinfo{year}{1998}\natexlab{}.
\newblock \showarticletitle{{OpenMP}: an industry standard {API} for
  shared-memory programming}.
\newblock \bibinfo{journal}{\emph{IEEE Computational Science and Engineering}}
  \bibinfo{volume}{5}, \bibinfo{number}{1} (\bibinfo{date}{Jan}
  \bibinfo{year}{1998}), \bibinfo{pages}{46--55}.
\newblock
\showISSN{1070-9924}


\bibitem[\protect\citeauthoryear{Davis and Hu}{Davis and Hu}{2011}]%
        {florida}
\bibfield{author}{\bibinfo{person}{Timothy~A. Davis} {and}
  \bibinfo{person}{Yifan Hu}.} \bibinfo{year}{2011}\natexlab{}.
\newblock \showarticletitle{The University of Florida Sparse Matrix
  Collection}.
\newblock \bibinfo{journal}{\emph{ACM Trans. Math. Softw.}}
  \bibinfo{volume}{38}, \bibinfo{number}{1}, Article \bibinfo{articleno}{1}
  (\bibinfo{date}{Dec.} \bibinfo{year}{2011}), \bibinfo{numpages}{25}~pages.
\newblock


\bibitem[\protect\citeauthoryear{Dongarra, Faverge, Ltaief, and
  Luszczek}{Dongarra et~al\mbox{.}}{2011}]%
        {Dongarra2011}
\bibfield{author}{\bibinfo{person}{Jack Dongarra}, \bibinfo{person}{Mathieu
  Faverge}, \bibinfo{person}{Hatem Ltaief}, {and} \bibinfo{person}{Piotr
  Luszczek}.} \bibinfo{year}{2011}\natexlab{}.
\newblock \showarticletitle{High Performance Matrix Inversion Based on {LU}
  Factorization for Multicore Architectures}. In
  \bibinfo{booktitle}{\emph{Proceedings of the 2011 ACM International Workshop
  on Many Task Computing on Grids and Supercomputers}}
  \emph{(\bibinfo{series}{MTAGS '11})}. \bibinfo{publisher}{ACM},
  \bibinfo{address}{New York, NY, USA}, \bibinfo{pages}{33--42}.
\newblock
\showISBNx{978-1-4503-1145-8}


\bibitem[\protect\citeauthoryear{Goedecker}{Goedecker}{1999}]%
        {Goedecker1999}
\bibfield{author}{\bibinfo{person}{Stefan Goedecker}.}
  \bibinfo{year}{1999}\natexlab{}.
\newblock \showarticletitle{Linear scaling electronic structure methods}.
\newblock \bibinfo{journal}{\emph{Reviews of Modern Physics}}
  \bibinfo{volume}{71} (\bibinfo{date}{Jul} \bibinfo{year}{1999}),
  \bibinfo{pages}{1085--1123}.
\newblock
Issue 4.


\bibitem[\protect\citeauthoryear{Higham and Lin}{Higham and Lin}{2011}]%
        {higham2011}
\bibfield{author}{\bibinfo{person}{Nicholas~J Higham} {and}
  \bibinfo{person}{Lijing Lin}.} \bibinfo{year}{2011}\natexlab{}.
\newblock \showarticletitle{A {Schur--Pad{\'e}} algorithm for fractional powers
  of a matrix}.
\newblock \bibinfo{journal}{\emph{SIAM J. Matrix Anal. Appl.}}
  \bibinfo{volume}{32}, \bibinfo{number}{3} (\bibinfo{year}{2011}),
  \bibinfo{pages}{1056--1078}.
\newblock


\bibitem[\protect\citeauthoryear{Higham and Lin}{Higham and Lin}{2013}]%
        {higham2013}
\bibfield{author}{\bibinfo{person}{Nicholas~J. Higham} {and}
  \bibinfo{person}{Lijing Lin}.} \bibinfo{year}{2013}\natexlab{}.
\newblock \showarticletitle{An Improved {Schur--Padé} Algorithm for Fractional
  Powers of a Matrix and Their {Fréchet} Derivatives}.
\newblock \bibinfo{journal}{\emph{SIAM J. Matrix Anal. Appl.}}
  \bibinfo{volume}{34}, \bibinfo{number}{3} (\bibinfo{year}{2013}),
  \bibinfo{pages}{1341--1360}.
\newblock


\bibitem[\protect\citeauthoryear{Intel}{Intel}{2017a}]%
        {IntelMKL}
\bibfield{author}{\bibinfo{person}{Intel}.} \bibinfo{year}{2017}\natexlab{a}.
\newblock \bibinfo{title}{{Math Kernel Library (MKL)}}.
\newblock
\newblock
\urldef\tempurl%
\url{http://www.intel.com/software/products/mkl/}
\showURL{%
\tempurl}


\bibitem[\protect\citeauthoryear{Intel}{Intel}{2017b}]%
        {IntelMPI}
\bibfield{author}{\bibinfo{person}{Intel}.} \bibinfo{year}{2017}\natexlab{b}.
\newblock \bibinfo{title}{{MPI Library}}.
\newblock
\newblock
\urldef\tempurl%
\url{https://software.intel.com/en-us/intel-mpi-library}
\showURL{%
\tempurl}


\bibitem[\protect\citeauthoryear{Klav{\'\i}k, Malossi, Bekas, and
  Curioni}{Klav{\'\i}k et~al\mbox{.}}{2014}]%
        {KlavikMalossiBekasEtAl2014}
\bibfield{author}{\bibinfo{person}{P. Klav{\'\i}k}, \bibinfo{person}{A.~C.~I.
  Malossi}, \bibinfo{person}{C. Bekas}, {and} \bibinfo{person}{A. Curioni}.}
  \bibinfo{year}{2014}\natexlab{}.
\newblock \showarticletitle{{Changing Computing Paradigms Towards Power
  Efficiency}}.
\newblock \bibinfo{journal}{\emph{Philosophical Transactions of the Royal
  Society {A}: Mathematical, Physical \& Engineering Sciences}}
  \bibinfo{volume}{372}, \bibinfo{number}{2018} (\bibinfo{year}{2014}).
\newblock


\bibitem[\protect\citeauthoryear{Kohn}{Kohn}{1999}]%
        {Kohn1999}
\bibfield{author}{\bibinfo{person}{W. Kohn}.} \bibinfo{year}{1999}\natexlab{}.
\newblock \showarticletitle{Nobel Lecture: Electronic structure of matter --
  wave functions and density functionals}.
\newblock \bibinfo{journal}{\emph{Reviews of Modern Physics}}
  \bibinfo{volume}{71} (\bibinfo{date}{Oct} \bibinfo{year}{1999}),
  \bibinfo{pages}{1253--1266}.
\newblock
Issue 5.


\bibitem[\protect\citeauthoryear{Kolberg, Bohlender, and Claudio}{Kolberg
  et~al\mbox{.}}{2008}]%
        {Kolberg2008}
\bibfield{author}{\bibinfo{person}{Mariana Kolberg}, \bibinfo{person}{Gerd
  Bohlender}, {and} \bibinfo{person}{Dalcidio Claudio}.}
  \bibinfo{year}{2008}\natexlab{}.
\newblock \bibinfo{booktitle}{\emph{Improving the Performance of a Verified
  Linear System Solver Using Optimized Libraries and Parallel Computation}}.
\newblock \bibinfo{publisher}{Springer Berlin Heidelberg},
  \bibinfo{address}{Berlin, Heidelberg}, \bibinfo{pages}{13--26}.
\newblock
\showISBNx{978-3-540-92859-1}


\bibitem[\protect\citeauthoryear{Lanczos}{Lanczos}{1950}]%
        {lanczos1950}
\bibfield{author}{\bibinfo{person}{C. Lanczos}.}
  \bibinfo{year}{1950}\natexlab{}.
\newblock \showarticletitle{An Iteration Method for the Solution of the
  Eigenvalue Problem of Linear Differential and Integral Operators}.
\newblock \bibinfo{journal}{\emph{Journal of research of the National Bureau of
  Standards}} \bibinfo{volume}{45}, \bibinfo{number}{4} (\bibinfo{date}{Oct.}
  \bibinfo{year}{1950}), \bibinfo{pages}{255--282}.
\newblock


\bibitem[\protect\citeauthoryear{Lass, Kühne, and Plessl}{Lass
  et~al\mbox{.}}{2017}]%
        {lass17_esl}
\bibfield{author}{\bibinfo{person}{Michael Lass}, \bibinfo{person}{Thomas~D.
  Kühne}, {and} \bibinfo{person}{Christian Plessl}.}
  \bibinfo{year}{2017}\natexlab{}.
\newblock \showarticletitle{Using Approximate Computing for the Calculation of
  Inverse Matrix p-th Roots}.
\newblock \bibinfo{journal}{\emph{IEEE Embedded Systems Letters}}
  (\bibinfo{year}{2017}).
\newblock
\showeprint[arxiv]{1703.02283}
\newblock
\shownote{Accepted for publication.}


\bibitem[\protect\citeauthoryear{Le~Gall}{Le~Gall}{2014}]%
        {LeGall2014}
\bibfield{author}{\bibinfo{person}{Fran\c{c}ois Le~Gall}.}
  \bibinfo{year}{2014}\natexlab{}.
\newblock \showarticletitle{Powers of Tensors and Fast Matrix Multiplication}.
  In \bibinfo{booktitle}{\emph{Proceedings of the 39th International Symposium
  on Symbolic and Algebraic Computation}} \emph{(\bibinfo{series}{ISSAC '14})}.
  \bibinfo{publisher}{ACM}, \bibinfo{address}{New York, NY, USA},
  \bibinfo{pages}{296--303}.
\newblock
\showISBNx{978-1-4503-2501-1}


\bibitem[\protect\citeauthoryear{Lehoucq, Sorensen, and Yang}{Lehoucq
  et~al\mbox{.}}{1998}]%
        {arpack}
\bibfield{author}{\bibinfo{person}{R.B. Lehoucq}, \bibinfo{person}{D.C.
  Sorensen}, {and} \bibinfo{person}{C. Yang}.} \bibinfo{year}{1998}\natexlab{}.
\newblock \bibinfo{booktitle}{\emph{ARPACK Users' Guide: Solution of
  Large-scale Eigenvalue Problems with Implicitly Restarted Arnoldi Methods}}.
\newblock \bibinfo{publisher}{Society for Industrial and Applied Mathematics}.
\newblock
\showISBNx{9780898714074}
\showLCCN{98061204}


\bibitem[\protect\citeauthoryear{Lin, Yang, Lu, Ying, and E}{Lin
  et~al\mbox{.}}{2011}]%
        {Lin2011}
\bibfield{author}{\bibinfo{person}{Lin Lin}, \bibinfo{person}{Chao Yang},
  \bibinfo{person}{Jianfeng Lu}, \bibinfo{person}{Lexing Ying}, {and}
  \bibinfo{person}{Weinan E}.} \bibinfo{year}{2011}\natexlab{}.
\newblock \showarticletitle{A Fast Parallel Algorithm for Selected Inversion of
  Structured Sparse Matrices with Application to {2D} Electronic Structure
  Calculations}.
\newblock \bibinfo{journal}{\emph{SIAM Journal on Scientific Computing}}
  \bibinfo{volume}{33}, \bibinfo{number}{3} (\bibinfo{date}{June}
  \bibinfo{year}{2011}), \bibinfo{pages}{1329--1351}.
\newblock
\showISSN{1064-8275}


\bibitem[\protect\citeauthoryear{Maschhoff and Sorensen}{Maschhoff and
  Sorensen}{1996}]%
        {p_arpack}
\bibfield{author}{\bibinfo{person}{Kristyn~J Maschhoff} {and}
  \bibinfo{person}{Danny~C Sorensen}.} \bibinfo{year}{1996}\natexlab{}.
\newblock \showarticletitle{{P\_ARPACK}: An efficient portable large scale
  eigenvalue package for distributed memory parallel architectures}. In
  \bibinfo{booktitle}{\emph{International Workshop on Applied Parallel
  Computing}}. Springer, \bibinfo{pages}{478--486}.
\newblock


\bibitem[\protect\citeauthoryear{{Message Passing Interface Forum}}{{Message
  Passing Interface Forum}}{1994}]%
        {MPI}
\bibfield{author}{\bibinfo{person}{{Message Passing Interface Forum}}.}
  \bibinfo{year}{1994}\natexlab{}.
\newblock \bibinfo{booktitle}{\emph{{MPI}: A Message-Passing Interface
  Standard}}.
\newblock \bibinfo{type}{{T}echnical {R}eport}.
  \bibinfo{institution}{University of Tennessee}, \bibinfo{address}{Knoxville,
  TN, USA}.
\newblock


\bibitem[\protect\citeauthoryear{Mohr, Dawson, Wagner, Caliste, Nakajima, and
  Genovese}{Mohr et~al\mbox{.}}{2017}]%
        {Mohr2017}
\bibfield{author}{\bibinfo{person}{Stephan Mohr}, \bibinfo{person}{William
  Dawson}, \bibinfo{person}{Michael Wagner}, \bibinfo{person}{Damien Caliste},
  \bibinfo{person}{Takahito Nakajima}, {and} \bibinfo{person}{Luigi Genovese}.}
  \bibinfo{year}{2017}\natexlab{}.
\newblock \showarticletitle{Efficient Computation of Sparse Matrix Functions
  for Large-Scale Electronic Structure Calculations: The {CheSS} Library}.
\newblock \bibinfo{journal}{\emph{Journal of Chemical Theory and Computation}}
  \bibinfo{volume}{13}, \bibinfo{number}{10} (\bibinfo{year}{2017}),
  \bibinfo{pages}{4684--4698}.
\newblock
\newblock
\shownote{PMID: 28873312.}


\bibitem[\protect\citeauthoryear{Mohr, Ratcliff, Boulanger, Genovese, Caliste,
  Deutsch, and Goedecker}{Mohr et~al\mbox{.}}{2014}]%
        {mohr2014}
\bibfield{author}{\bibinfo{person}{S. Mohr}, \bibinfo{person}{L.E. Ratcliff},
  \bibinfo{person}{P. Boulanger}, \bibinfo{person}{L. Genovese},
  \bibinfo{person}{D. Caliste}, \bibinfo{person}{T. Deutsch}, {and}
  \bibinfo{person}{S. Goedecker}.} \bibinfo{year}{2014}\natexlab{}.
\newblock \showarticletitle{Daubechies wavelets for linear scaling density
  functional theory}.
\newblock \bibinfo{journal}{\emph{J Chem Phys}} \bibinfo{volume}{140},
  \bibinfo{number}{20} (\bibinfo{year}{2014}), \bibinfo{pages}{204110}.
\newblock


\bibitem[\protect\citeauthoryear{Prodan and Kohn}{Prodan and Kohn}{2005}]%
        {prodan2005}
\bibfield{author}{\bibinfo{person}{E. Prodan} {and} \bibinfo{person}{W. Kohn}.}
  \bibinfo{year}{2005}\natexlab{}.
\newblock \showarticletitle{Nearsightedness of electronic matter}.
\newblock \bibinfo{journal}{\emph{Proceedings of the National academy of
  Sciences of the United States of America}} \bibinfo{volume}{102},
  \bibinfo{number}{33} (\bibinfo{year}{2005}), \bibinfo{pages}{11635--11638}.
\newblock


\bibitem[\protect\citeauthoryear{Richters and K\"uhne}{Richters and
  K\"uhne}{2014}]%
        {Richters2014}
\bibfield{author}{\bibinfo{person}{D. Richters} {and} \bibinfo{person}{T.~D.
  K\"uhne}.} \bibinfo{year}{2014}\natexlab{}.
\newblock \showarticletitle{Self-consistent field theory based molecular
  dynamics with linear system-size scaling}.
\newblock \bibinfo{journal}{\emph{Journal of Chemical Physics}}
  \bibinfo{volume}{140}, \bibinfo{number}{13} (\bibinfo{year}{2014}),
  \bibinfo{pages}{134109}.
\newblock

\newpage
\bibitem[\protect\citeauthoryear{Richters, Lass, Walther, Plessl, and
  Kühne}{Richters et~al\mbox{.}}{2018}]%
        {richters17_arxiv}
\bibfield{author}{\bibinfo{person}{Dorothee Richters}, \bibinfo{person}{Michael
  Lass}, \bibinfo{person}{Andrea Walther}, \bibinfo{person}{Christian Plessl},
  {and} \bibinfo{person}{Thomas~D. Kühne}.} \bibinfo{year}{2018}\natexlab{}.
\newblock \showarticletitle{A General Algorithm to Calculate the Inverse
  Principal $p$-th Root of Symmetric Positive Definite Matrices}.
\newblock \bibinfo{journal}{\emph{Communications in Computational Physics}}
  (\bibinfo{year}{2018}).
\newblock
\showeprint[arxiv]{1703.02456}
\newblock
\shownote{Accepted for publication.}


\bibitem[\protect\citeauthoryear{Schulz}{Schulz}{1933}]%
        {schulz1933}
\bibfield{author}{\bibinfo{person}{G. Schulz}.}
  \bibinfo{year}{1933}\natexlab{}.
\newblock \showarticletitle{{Iterative Berechung der reziproken Matrix}}.
\newblock \bibinfo{journal}{\emph{{ZAMM} - Zeitschrift fur Angewandte
  Mathematik und Mechanik}} (\bibinfo{year}{1933}).
\newblock


\bibitem[\protect\citeauthoryear{Shen}{Shen}{2006}]%
        {Shen2006}
\bibfield{author}{\bibinfo{person}{Kai Shen}.} \bibinfo{year}{2006}\natexlab{}.
\newblock \showarticletitle{Parallel Sparse {LU} Factorization on Different
  Message Passing Platforms}.
\newblock \bibinfo{journal}{\emph{J. Parallel and Distrib. Comput.}}
  \bibinfo{volume}{66}, \bibinfo{number}{11} (\bibinfo{date}{Nov.}
  \bibinfo{year}{2006}), \bibinfo{pages}{1387--1403}.
\newblock
\showISSN{0743-7315}


\bibitem[\protect\citeauthoryear{van~der Stappen, Bisseling, and van~de
  Vorst}{van~der Stappen et~al\mbox{.}}{1993}]%
        {vanderStappen1993}
\bibfield{author}{\bibinfo{person}{A.~Frank van~der Stappen},
  \bibinfo{person}{Rob~H. Bisseling}, {and} \bibinfo{person}{Johannes~G.G.
  van~de Vorst}.} \bibinfo{year}{1993}\natexlab{}.
\newblock \showarticletitle{Parallel Sparse {LU} Decomposition on a Mesh
  Network of Transputers}.
\newblock \bibinfo{journal}{\emph{SIAM J. Matrix Anal. Appl.}}
  \bibinfo{volume}{14}, \bibinfo{number}{3} (\bibinfo{year}{1993}),
  \bibinfo{pages}{853--879}.
\newblock


\end{thebibliography}

\end{document}